\documentclass[aps,preprint,a4paper,showpacs,amsmath,amssymb,floatfix,prl]{revtex4}
\usepackage{graphicx}
\DeclareGraphicsExtensions{.pdf}
\newcommand{\Op}[1]{{\bf {\hat {#1}}}}

\begin{document}

\title{Directional Quantum-Controlled Chemistry: Generating Aligned Ultracold Molecules via Photoassociation}
\author{S.\ Kallush$^{1,2}$, J.L. Carini$^{3\footnote{Present address: 400 Main St., East Hartford, CT
06118, USA}}$, P.L. Gould$^{3}$ and R. Kosloff$^{2}$}
\affiliation{$^1$ Department of Physics and Optical Engineering, ORT-Braude College, P.O. Box 78, 21982 Karmiel, Israel
\\$^2$The Fritz Haber Research Center, The Hebrew University of Jerusalem,
Jerusalem 91904, Israel
\\$^3$Department of Physics, University of Connecticut, Storrs, Connecticut 06269, USA}

\begin{abstract}
Photoassociation of ultracold atoms is shown to lead to alignment of the product molecules along the excitation laser polarization axis. We theoretically investigate pulsed photoassociation of $^{87}Rb$ atoms into a specific weakly-bound level of the a $^3\Sigma_u^+$ metastable electronic state and find both stationary and time-dependent field-free alignment. Although a transform-limited pulse yields significant alignment, a frequency-chirped pulse dramatically enhances the molecular formation rate at the cost of a slight decrease in the alignment. Employing multiple pulses synchronized with the vibrational and rotational periods leads to coherent enhancement of both population and alignment of the target state.

\end{abstract}

\pacs{32.80.Qk, 34.50.Rk,37.10.Mn,}

\maketitle

\section{Introduction}
Quantum coherent control was conceived as a method to direct chemical reactions to a desirable outcome \cite{tannor1985control,k46,brumer1986control}.
Despite the remarkable success of applying these techniques to induce unimolecular encounters such as 
photoisomerization\cite{prokhorenko2006coherent} and
photodissociation\cite{shapiro2003principles}, the raison d'\^etre for coherent control in the context of chemistry, 
namely, the assembling of multiple atoms or molecules into a new chemical species, was not realized until recently \cite{k297,k300}. 
The main obstacle has been the pre-entanglement of the reactants \cite{abrashkevich1998coherent}.
The approach to overcome this issue differs for hot and cold collisions.
For reactions in the high temperature regime  ($T \gg 1K$), one has to first distill entangled pairs out of initial thermal incoherent mixture \cite{k285}. 
In the cold ($T \ll 1K$) regime, quantum phenomena close to threshold result in the low angular momentum states being pre-entangled.
For the ultracold reactions, technological developments that allow pulse shaping in the time domain \cite{rogers2007} were used to demonstrate coherent control in ultracold
photoassociation \cite{k300}.

In this paper we present a new aspect of quantum control unique to photochemical reactions. 
Such control is able to align the products of the reaction along a well-defined direction in the lab frame. 
The discipline of lab-frame spectroscopy of molecules is currently attracting great attention. It has emerged as a crucial element in harmonic generation\cite{kanai2005quantum,mcfarland2008high,kraus2012}, directional molecular ionization/dissociation\cite{de2009,litvinyuk2003alignment} and other applications\cite{stapelfeldt2003,lemeshko2013,fleischer2012}. 
We employ here the concepts that are used to achieve alignment of already-existing molecules and adapt them into the framework of photoassociative molecule formation. Novel features arising in the system under study include the nanosecond timescale of the dynamics, and the adequacy of low intensities due to the resonant nature of the interaction. Although for the sake of simplicity, the demonstration presented here is for ultracold samples, in principle it could be extended to any thermal regime.

\section{The model}
\label{sec:model}

The model describes photoassociation (PA) of two Rb atoms to synthesize an aligned Rb$_2$ molecule. An ensemble 
of ultracold $^{87}Rb$ atoms is subjected to a 
time-dependent electric field linearly polarized along the $Z$ axis in the lab frame. The shaped laser pulse is employed to associate a pair of free atoms colliding on the ground electronic state $5S_{1/2} +5S_{1/2}$ asymptote into a bound state of the $0^{-}_g$ electronic state below the $5S_{1/2} + 5P_{3/2}$ asymptote.
Subsequently, but within the same pulse, a coherent coupling drives the excited molecule into a bound level of the $a^3\Sigma_u^+$ metastable state. 
This model extends the purely vibration-electronic system described in  \cite{k300,k308} by 
including the rotational degree of freedom. 

The intermediate excited vibrational state and the pulse parameters can be optimized to yield production rates up to $10^8$ molecules/s \cite{k308}. 
The $v'=31$ level is chosen here as the intermediate state, therefore the dynamics are described by the three rotational manifolds which correspond to (1) the scattering continuum of the ground electronic state, (2) the $0^{-}_{g}$ ($v'=31$) intermediate electronic state, and finally (3) the next-to-last $v''=39$ vibrational manifold of $a^3\Sigma_u^+$. 
For all the states involved in the process, the main contribution to the excitation/de-excitation process occurs at sufficiently large internuclear distance (around $R\approx 55$ bohrs) that Hund's case (c) is applicable. 

Under this assumption, the states are given by $ \left| n J M \Omega \right\rangle$, where $n$ stands for all the non-rotational quantum numbers, $J$ is the total angular momentum, $M$ is the projection of $J$ oמ the lab $Z$ axis, and $\Omega$ is the projection of $J$ on the molecule-fixed axis. 
Note that within Hund's case (c), the electronic orbital and spin angular momenta are not coupled to the molecular axis and therefore their projections are not separately conserved. Energies are measured relative to that of the scattering state. The Hamiltonian for the light-matter interaction has the form:
\begin{equation}
\Op{H}(t) = \Op{H}_0 +{\bf{\hat{\mu}}}\cdot{\bf{E}}(t) = \Op{H}_0 +\hat{\mu}E_{Z}(t)cos(\theta).
\end{equation}
where $\theta$ is the angle between the $Z$ axis and the molecular axis. The bound states energies of ro-vibronic levels within the field-free Hamiltonian $\Op{H}_0$ are given by $E_{nJ\Omega} = E_n + B_v(J(J+1)-\Omega^2)$, where $E_n$ is the vibronic energy and $B_v = \hbar^2/2m\left\langle R^2\right\rangle$ is the level's rotational constant computed from the nuclear wavefunction. Due to the large outer turning points, $B_v/\hbar$ is quite small, in the 10 $MHz$ range.  

The time-dependent electric field pulse is described as:
\begin{equation}
{\bf{E}}(t) = \hat{Z} E_0 \exp\left(-\left(\sigma^{-2}+i\chi\right)t^2\right)
\end{equation}
where $E_0$ is the field amplitude, $\sigma$ is the temporal pulse width, and $\chi$ is the linear chirp rate.
For linearly-polarized light, the rotational part of the transition dipole moment between the different manifolds is given by:
\begin{equation}
\left\langle n'J'M\Omega' |\Op{\mu}| nJM\Omega\right\rangle	= F_{nJ\Omega}^{n'J'\Omega'}\sqrt{(2J+1)(2J'+1)}\left(-1\right)^{M-\Omega}\left(  {\begin{array}{ccc}
   J' & 1 & J \\ M & 0 & M \ \end{array} }\right)   \left(  {\begin{array}{ccc}
   J' & 1 & J \\ \Omega' & q & \Omega \ \end{array} }\right)    
\end{equation}
where common notation is used for the Wigner 3-j symbols, $q = -1,0,1$.
$F_{nJ\Omega}^{n'J'\Omega'}$ are Franck-Condon (FC) factors between two vibrational levels, and $cos(\theta) \propto D^1_{00}$, where $D^J_{M\Omega}$ is the rotational matrix wavefunction\cite{zare}.  
The FC factors are obtained using  nuclear wavefunctions computed by a mapped Fourier grid \cite{k145,k224}. 
The energy eigenfunctions are obtained by diagonalizing the time-independent nuclear Hamiltonian 
$\Op{H}_{nJ\Omega} = \Op{T}+\Op{V}_{nJ\Omega}$ where $\Op{T}$ is the kinetic energy operator and the potential $\Op{V}_{nJ\Omega}$ reads: 
\begin{equation}
	\Op{V}_{nJ\Omega} = V_n(R) + \frac{\hbar^2\left(J(J+1)-\Omega^2\right)}{2mR^2}~~,
\end{equation}
where $V_n$ are the Born-Oppenheimer potential curves. 

The final observables such as population and alignment are obtained by solving the time-dependent Schr\"odinger equation for each of the initial thermal states in the scattering manifold. 
The observables are computed by averaging  over an incoherent sum of the individual runs. 
Note that due to the nuclear spin statistics, the odd and even $J$ states lead to symmetries that have to be weighted according to: $P_{odd}/P_{even} = (I+1)/I$, where $I=3/2$ is the nuclear spin of $^{87}Rb$. 

The timescales of the dynamics are comparable to the spontaneous decay time ($22$ ns). 
As was shown in Ref.\cite{k308}, the incoherent part of the dynamics has a negligible effect on the overall yield. 
In addition, as alignment can emerge only from coherent processes, spontaneous decay cannot contribute.  

\section{Molecular alignment}
\label{sec:results}

\begin{figure}[tbp] 
 \centerline{
    \mbox{\includegraphics[width=5.00in]{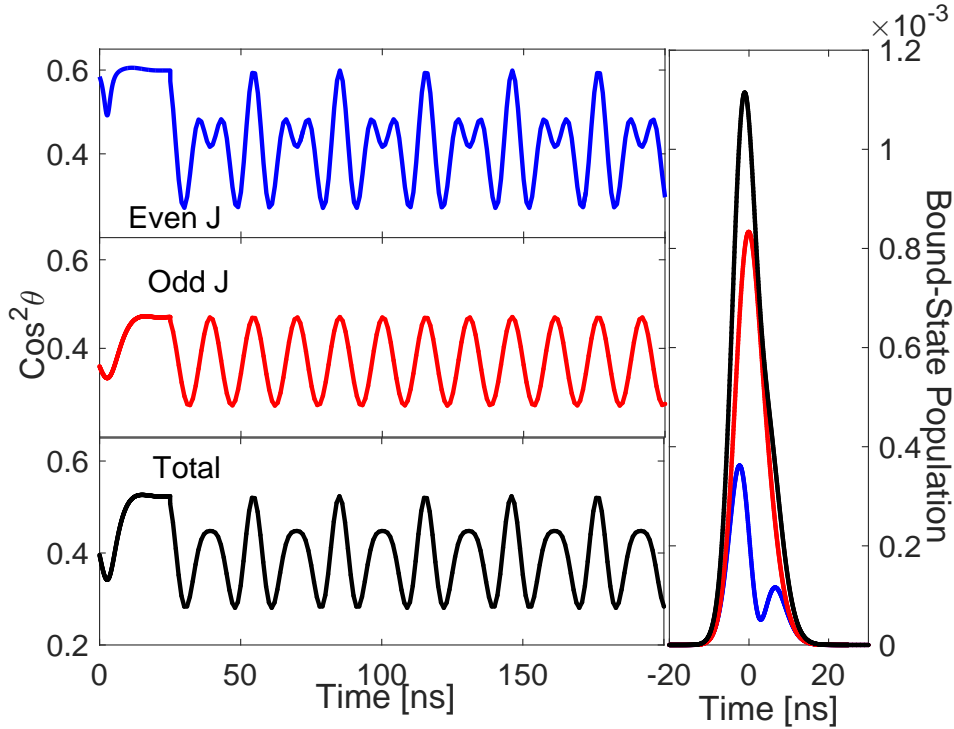}}}
\caption{(Color online) Alignment of the photoassociated molecules: (Main) Alignment vs. time for the even (top) and odd (middle) $J$ states and total population (bottom). (Right) Bound-state population vs. time for the even (blue) and odd (red) $J$ states, and the total (black). The pulse is a $\sigma = 10$ ns transform-limited pulse, with peak intensity $1000~ W/cm^2$ and a detuning $\delta = 0$.} 
\label{fig1}
\end{figure}
The molecular alignment after PA by a single  transform-limited (TL) pulse, quantified by the expectation value of $\langle cos^2\theta \rangle$, is shown in Fig. \ref{fig1}. The right inset panel presents the temporal bound-state population of the various symmetries and of the total weighted sum during the pulse. For an isotropic ensemble in thermal equilibrium, all  $M$ states for any given $J$ are equally populated, so the probabilities to be aligned to any arbitrary direction are equal, and $\langle cos^2\theta \rangle = 1/3$.

Two types of angular anisotropy are visible in Figure \ref{fig1}: 
(1) The time-average of the alignment is not 1/3. This indicates a stationary alignment, which results from the non-thermal population for different $M$ states.
(2) The alignment oscillates in time. This phenomena emerges from coherences between different $J$ states 
with the same M value. The periodicity of the signal is found to be $30.7~ns = 1/2B$, where $B$ is the rotational constant of $a^3\Sigma_u^+\left( v'' = 39\right)$. 
This typical lab-frame alignment is the starting point of many experiments
in the context of lab-frame spectroscopy \cite{stapelfeldt2003}.

\begin{figure}[tbp] 
 \centerline{
    \mbox{\includegraphics[width=7.00in]{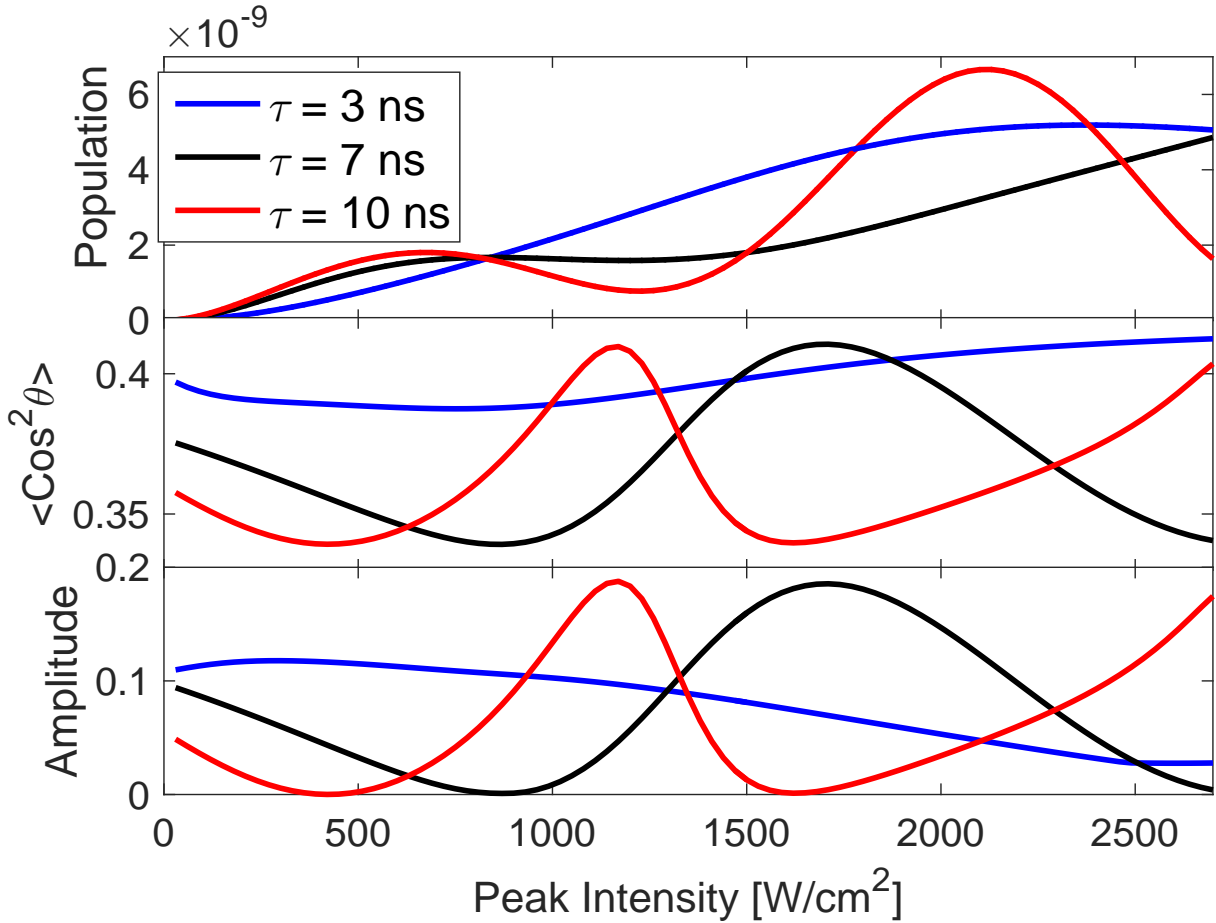}}}
\caption{(Color online) Transform-limited pulses: (upper) the bound-state population, (middle) the stationary alignment, and (lower) the amplitude of the dynamic alignment, as functions of the peak intensity. The three curves correspond to different pulse widths.} 
\label{fig2}
\end{figure}
Figure \ref{fig2} shows the influence of peak intensity and pulse width on the alignment and PA yield.
The bound-state populations are similar to those in our previous study \cite{k300} for TL pulses at these intensities and pulse widths. 
The pulses with widths $\sigma = 7 ~ns$ and $\sigma=10 ~ns$  show a similar dependency of the stationary and dynamic alignment on intensity. 
The shorter $\sigma = 3$ ns pulse shows a different intensity dependence. Note that the binding energy of the $v''=39$ state is $764 ~ MHz$, which corresponds to a period of $1.3$ ns. Hence, for a longer pulse, with narrow frequency bandwidth, the static and dynamic alignments  emerge from the same origin, the population of the same single state, and therefore have the same behavior. This is not the case for shorter pulses, or alternatively for a chirped pulse (see below). The alignment presented in Fig. \ref{fig2} under almost all of the conditions is very significant. 

The field that drives the process has a well-defined polarization direction. Alignments of this magnitude should be detectable with resonable experimental resolution. 
The maximum (aligned) and minimum (anti-aligned) values with the optimal parameters reach 0.6 and 0.2, respectively. These values depend on the number of $M$ values that contribute to the final superposition.
For ultracold PA they correspond to low $J$ cases  and $M = 0 (\pm 1)$ for the most aligned (anti-aligned).  
For bulk gas-phase molecules, increasing the pulse intensity can add more $M$ states to the superposition, leading to higher alignment. 
However, under ultracold initial conditions, the maximum $J$ values achievable during PA are limited by rotational barriers and the resulting extremely sharp exponential decay of the Franck-Condon factors. 
\begin{figure}[tbp] 
 \centerline{
    \mbox{\includegraphics[width=7.00in]{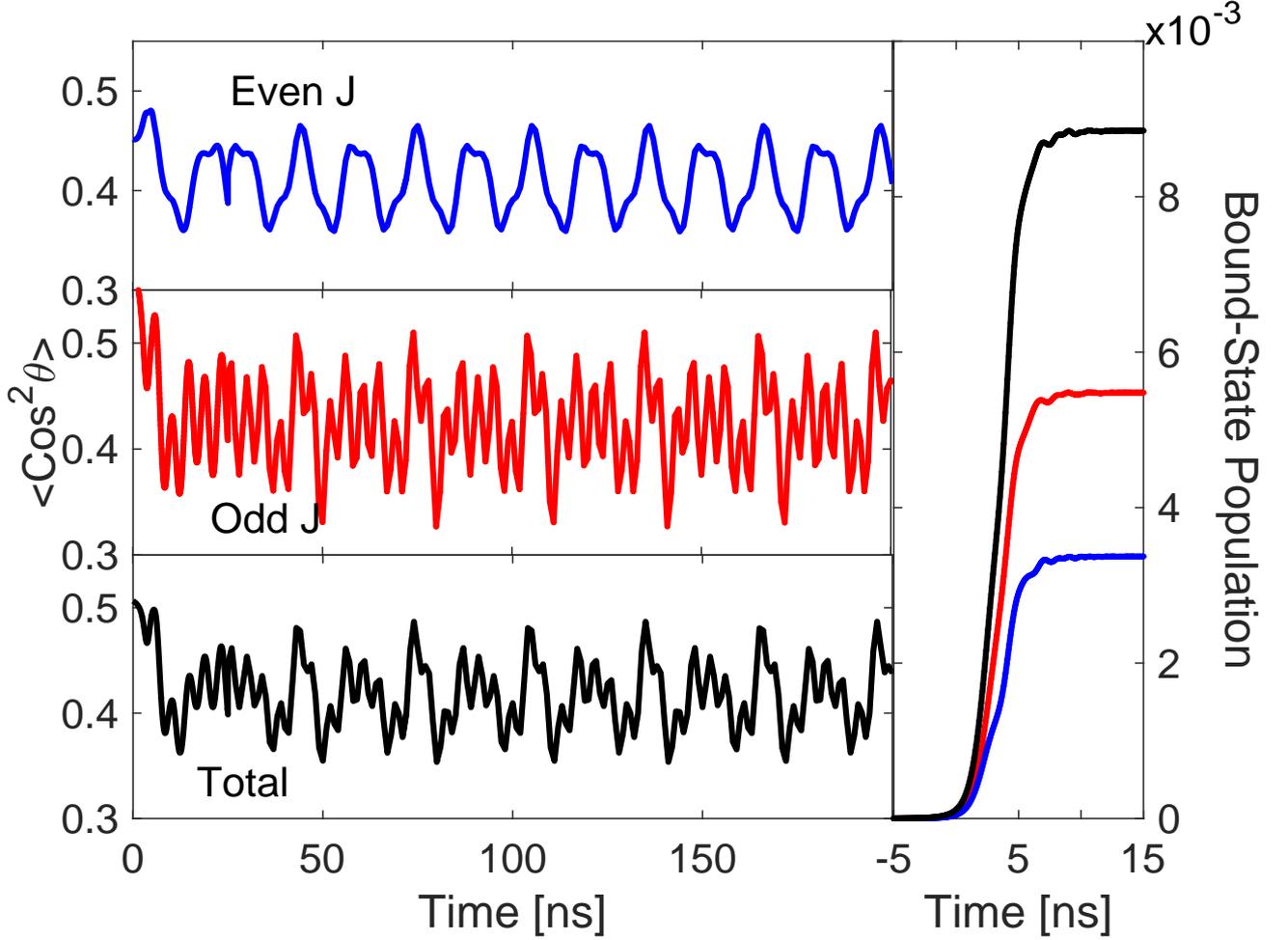}}}
\caption{(Color online) Chirped pulses: Similar to Fig. \ref{fig1}, with the same pulse width and peak intensity. The pulse is chirped at a rate of $100 ~ MHz/ns$, and is on resonance with $v'=31$ at the peak of the pulse.} 
\label{fig3}
\end{figure}

Figure \ref{fig3} shows the alignment and population in the target state with a positively chirped pulse with a chirp rate $\chi = 100 ~ MHz/ns$. As expected \cite{k300}, the bound-state population here is dramatically increased, a result of the the light becoming resonant with the downward transition to the bound manifold later in the pulse. Being in resonance, the population of the target bound state is maintained and not simply a transient as in the TL case in Fig. \ref{fig1}. Consequently, the majority of the population arrives into these states only during the later part of the pulse. The alignment here is also quite pronounced and contains the same period of $30.7$ ns, but due to the population of many rotational states by the chirp, the signal here also has faster oscillations. 

\begin{figure}[tbp] 
 \centerline{
    \mbox{\includegraphics[width=7.00in]{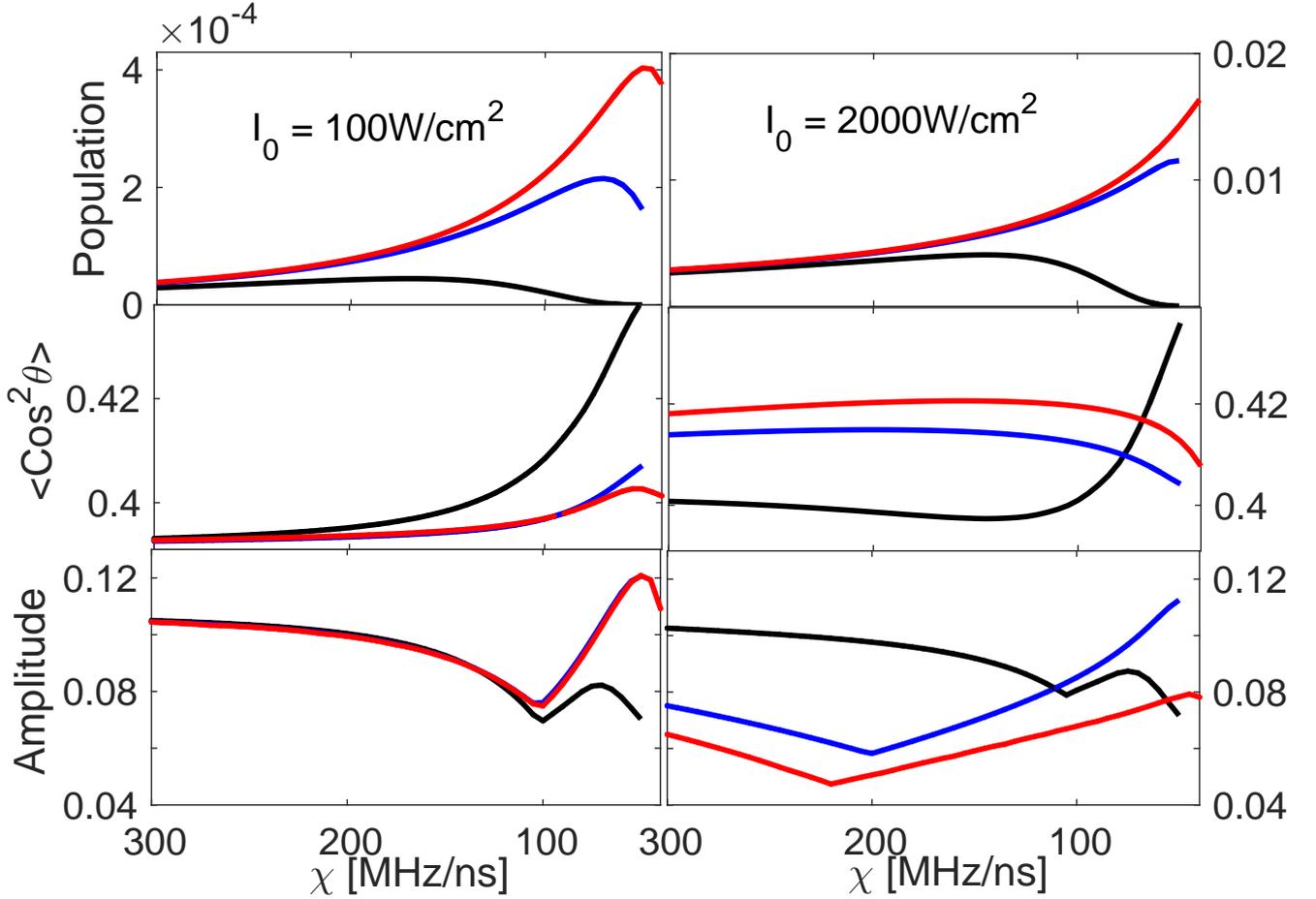}}}
\caption{(Color online) Photoassociation by chirped pulses: (upper) the bound-state population, (middle) the stationary alignment, and (lower) the amplitude of the dynamic alignment as functions of the chirp rate. Left panels correspond to a peak intensity of $I_0 = 100~W/cm^2$, and right panels to $I_0 = 2000~W/cm^2$. The three curves correspond to the different pulse widths of Fig \ref{fig2}.} 
\label{fig4}
\end{figure}
We have found \cite{k308} that the PA yield can be optimized by the appropriate choice of chirp parameters. The bound-state population and alignment vs. chirp rate for two intensities and three pulse widths are shown in Fig. \ref{fig4}. The optimization of bound-state population is due to a competition between two factors. One the one hand, a slow chirp is more adiabatic and therefore more efficient. On the other hand, for a fixed pulse width, if the chirp is too slow, the resonance for the de-excitation step is reached late in the pulse, when the intensity is low. This results in a low post-pulse bound-state population, similar to the TL values shown in Fig. \ref{fig2}. As expected, the longer the pulse, the slower the optimizing chirp.

Due to the fact that all of the states pass through resonance at some time during the pulse, the peak intensity itself plays a minor role in the dependence of the final population on chirp rate. 
Because of the similar outer turning points of the two bound states, the de-excitation step in the process is easily saturated, and the bound-state population scales linearly with the field intensity. 
This relatively simple behavior is not maintained for the static and dynamic alignment: For low intensities the alignment is nearly the same for pulse widths above 5 ns. 
Once the states that are relevant for the alignment are reached during the chirp, the relative fractions of the population and their coherences are not changed, and the alignment does not depend on the pulse width. For higher intensities, interference effects start to play a role and the alignments vary significantly with respect to the various parameters. Note that due to the relatively large number of states that participate for the chirped pulses, the final static and dynamic alignment values are moderately smaller than those achieved with TL pulses. This inferior performance could probably be improved with more sophisticated manipulation of the temporal phase of the field, e.g., with the use of some formal control scheme such as local or optimal control.

\begin{figure}[tbp] 
 \centerline{
    \mbox{\includegraphics[width=7.00in]{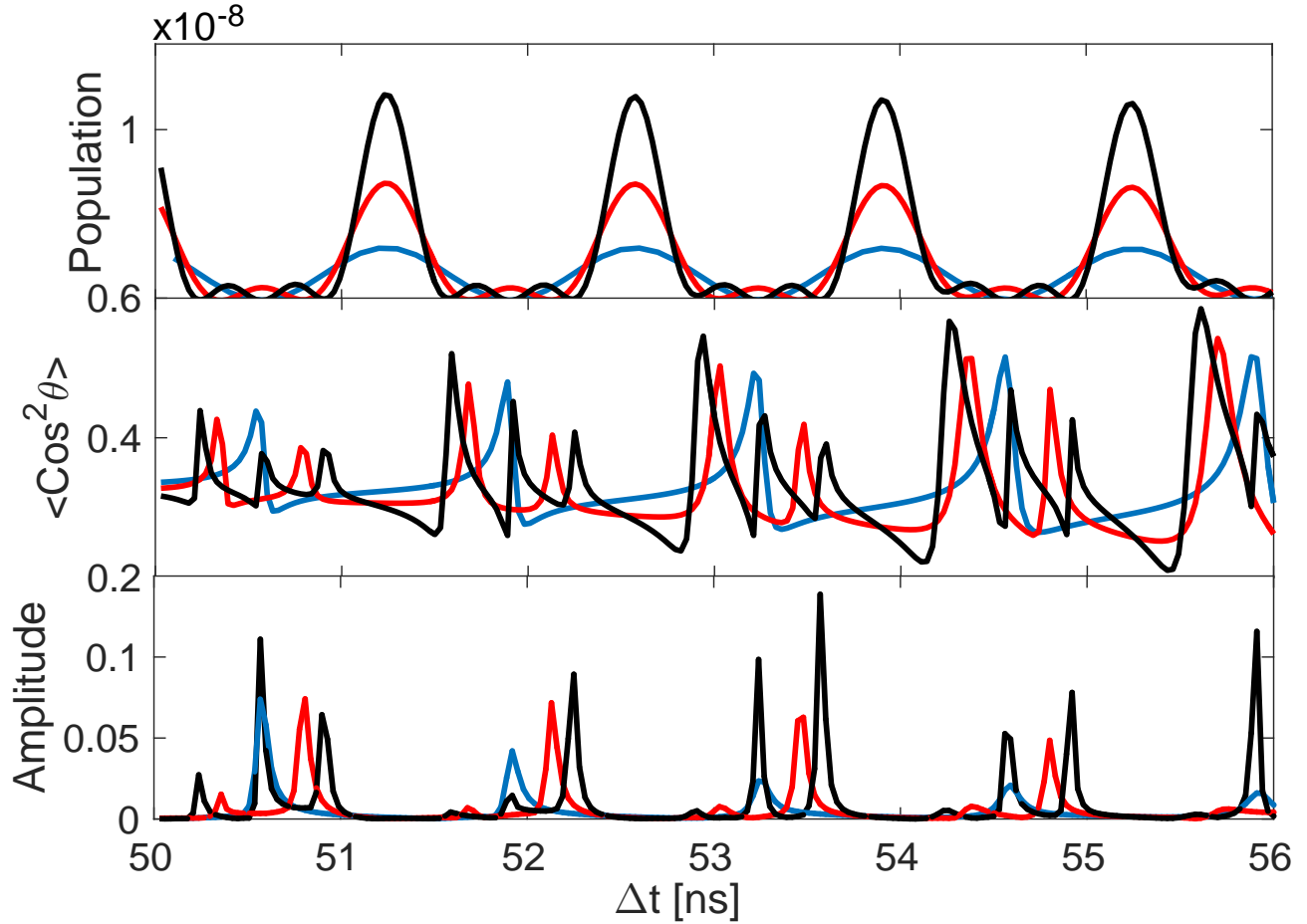}}}
\caption{(Color online) Photoassociation by pulse trains: (upper) the bound-state population, (middle) the stationary alignment, and (lower) the amplitude of the dynamic alignment as functions of the delay between the pulses. The three curves at each panel are for 2 (blue), 3 (red), and 4 (black) pulses. All the pulses are $10 ~ ns$ TL with $I_0 = 500 ~ W/cm^2$.} 
\label{fig5}
\end{figure}
In cold PA of dilute samples, coherences within a given electronic state are long lasting.
As a result, a series of photoassociation pulses will add coherently to the process.
The response to multiple pulses is examined in Fig. \ref{fig5}. The use of pulse trains is common in both ultracold PA \cite{k280} and lab-frame spectroscopy \cite{Bucksbaum2015}. The figure displays the bound-state population, the time-averaged alignment and the dynamic alignment amplitude vs. the time delay between the pulses. The three curves correspond to the response after two, three and four pulses. To avoid any overlap between the pulses, the minimum delay is taken at be $5\sigma = 50 $ ns. Only delays up to $56 ~ ns$ are shown in the plot. For longer time delays see Fig. 1 in the supplementary material of this paper. 

Two distinct oscillation periods are visible in the graphs: (I) The faster $\tau \sim 1.3 ~ns$ period relates to the vibrational binding energy of the bound state. Note that in contrast to the conventional pulse trains that are applied to align molecules in bulk media, here newly aligned molecules are added coherently at each pulse. 
Thus there is a phase relation between the molecules that were created with different pulses, and interference takes place.
(II) The slower time period $\tau \sim 61 ~ns$ (visible more clearly in the supplementary figure) 
relates to the slower rotational energy gap of 1/B between the rotational energy states in the manifold. 
Note that due to the fact that the two periods are by no means commensurate, quasiperiodic dynamics occurs.

The interference pattern of the bound-state population for n pulses resembles, for short pulse delays, the intensity pattern of n-slit interference. The patterns of the static and dynamic alignments (middle and bottom panels) show more intriguing effects: When $n$ pulses are applied, the fundamental signal obtained for 2 pulses is maintained, but each peak is split into a multiplet of $n-1$ peaks whose locations are approximately symmetric about the original $n=2$ peak. Comparing the static alignment obtained after single and multiple pulses shows that the extreme values of $0.6~ (0.2)$ for the static alignment (anti-alignment) are obtained only under multiple-pulse conditions, taking advantage of the instantaneous phase between the bound and scattering manifolds to create interferences.
 	
\section{Discussion and Summary} 
\label{sec:discussion}

Molecular alignment of photoassociated products has never been measured directly. The prerequisite is a PA pulse which is shorter than the rotational period.
The alignment of magnetoassociated ultracold molecules with respect to the magnetic field was demonstrated and examined recently\cite{Denschlag}. A detection method should be sensitive to the alignment under the specific conditions of dilute ultracold samples.
Birefringence, a common method of measuring the alignment in a bulk medium, is not appropriate due to the relatively small density of photoassociated molecules. Other approaches based on multiphoton excitation, such as Coulomb explosion \cite{Vrakking} and second-harmonic generation \cite{spector} could be employed. A promising method is to extend resonance-enhanced multiphoton ionization (REMPI) that has been used for conventional PA \cite{k300} in a fashion similar to \cite{Ohshima}. A pulse duration of a few nanoseconds should provide adequate resolution.
A linearly-polarized REMPI pulse can be tuned to an electronic $\Pi$ state to induce electronic excitations perpendicular to the molecule axis. 
For molecules which are aligned to the lab-fixed axis, a difference in the molecular signal for REMPI pulse polarizations parallel and perpendicular to the PA polarization should give a clear 
indication of the alignment. For the alignment values presented in the current work, we can estimate the ratio of the signals to be at most on the order of $S_{\bot}/S_{||} \sim 3$. 
After the PA generates aligned $v"=39$ molecules, additional alignment can be obtained by excitation/de-excitation cycles into lower vibrational states.

The introduction of lab-frame spectroscopy into coherently-controlled ultracold reactions raises several interesting applications. For heteronuclear ultracold molecules that possess a permanent dipole moment, controlled PA can generate macroscopic orientation of the whole ensemble. Taking advantage of the timing, this orientation can be field free. Furthermore, under BEC conditions, the induced macroscopic dipole could exhibit phenomena that originate in the quantum character of the molecules but give signatures on the macroscopic scale. A concerted rotation of the quantum condensed phase can lead to effects which will have analogs in other classical-like condensed-phase dynamics. 

\begin{acknowledgments}
This work is supported by the National Science Foundation through grant number PHY-1506244 and the US-Israel Binational Science Foundation through grant number 2012021.

\end{acknowledgments}
\bibliography{ref}

\bibliographystyle{unsrt}

\end{document}